\documentclass[pre,twocolumn,showpacs,preprintnumbers,amsmath,amssymb]{revtex4}
\usepackage{graphics}

\begin{document}

\title{Unrestricted Hartree-Fock theory of Wigner crystals}

\author{J. R. Trail}
\email{jrt32@cam.ac.uk}
\affiliation{TCM Group, Cavendish Laboratory, University of Cambridge,
 Madingley Road, Cambridge, CB3 0HE, UK}
\author{M. D. Towler}
\affiliation{TCM Group, Cavendish Laboratory, University of Cambridge,
 Madingley Road, Cambridge, CB3 0HE, UK}
\author{R. J. Needs}
\affiliation{TCM Group, Cavendish Laboratory, University of Cambridge,
 Madingley Road, Cambridge, CB3 0HE, UK}

\date{April, 2003}

\begin{abstract}
We demonstrate that unrestricted Hartree-Fock theory applied to
electrons in a uniform potential has stable Wigner crystal solutions
for $r_s \geq 1.44$ in two dimensions and $r_s \geq 4.5$ in three
dimensions.  The correlation energies of the Wigner crystal phases are
considerably smaller than those of the fluid phases at the same
density.
\end{abstract}

\pacs{71.10.Ca, 71.15.Nc, 73.20.Qt}

\maketitle

Wigner~\cite{wigner} first predicted that a system of electrons in a
uniform potential would crystallize at low densities.  Localizing
electrons around lattice sites increases their kinetic energy, but at
sufficiently low densities the reduction in interaction energy is
always greater.  The Wigner crystal remains a theoretical prediction
in three dimensions (3D), but in two dimensions (2D) Wigner crystals
have been created on a liquid helium surface~\cite{grimes} and at the
interface between two semiconductors~\cite{andrei,willett}.  It has
been suggested that electrons forming a Wigner crystal might
eventually be used as quantum bits (qubits) in a quantum
computer~\cite{platzman}.

The widely-studied model system of electrons in a uniform potential
has yielded many insights into electronic many-body phenomena.  The
most accurate calculations performed to date for the zero temperature
ground state phases of this system have used the diffusion quantum
Monte Carlo (DMC) method~\cite{Cep&Ald,rmp}.  A Wigner crystal may
also be described as a vibrating lattice of electrons.  When harmonic
phonon vibrations and anharmonic terms are included the resulting
energies are very similar to DMC ones~\cite{carr}.  A recent
Hartree-Fock study of small numbers of electrons confined by an
external potential revealed a transition from a Fermi fluid to a
Wigner molecule state~\cite{yannouleas}.  In this paper we also employ
the Hartree-Fock approximation which gives a description of Wigner
crystals in terms of Einstein oscillators, but including anharmonic
and exchange effects.

Within Hartree-Fock theory the paramagnetic (unpolarized) fluid phase
is unstable to the ferromagnetic (fully polarized) fluid for values of
the density parameter, $r_s$~\cite{density}, greater than 2.01 in 2D,
and greater than 5.45 in 3D.  Hartree-Fock theory also predicts that
the paramagnetic fluid is unstable to the formation of a spin density
wave~\cite{overhauser}.  The introduction of electron correlation
changes the picture dramatically, with the instability of the
paramagnetic to the ferromagnetic fluid being shifted to $r_s \simeq
26$ in 2D~\cite{attaccalite_2002} while in 3D a second order
transition to a partially polarized fluid is predicted to occur at
$r_s \simeq 50$~\cite{zong_2002}.  The spin-density-wave instability
may be entirely eliminated.  DMC calculations also predict the
occurrence of Wigner crystal phases for $r_s >
35$~\cite{rapisarda,tanatar} in 2D and $r_s > 65-100$ in
3D~\cite{Cep&Ald,ortiz_1999}.

Within a mean-field theory of electron systems the interactions are
replaced by a potential which acts on each electron orbital
separately.  The wave function is then simply a determinant of single
particle orbitals.  Two criteria are required for a mean field
description of Wigner crystals.  In a Wigner crystal the electrons are
localized individually, not in up and down spin pairs.  To describe
this situation we must use a mean field theory in which the potentials
felt by the up and down spin orbitals are different.  The second issue
is that some theories, such as standard implementations of density
functional theory, suffer from a spurious effect whereby the electrons
interact with themselves, giving a ``self-interaction'' error.  The
energy lowering on crystallization derives from the reduction of the
interaction energy by spatially separating the electrons.  A theory
which suffers from self interaction will tend to overestimate the
interaction energy of separated electrons, destabilizing Wigner
crystals.\cite{yannouleas} The unrestricted Hartree-Fock theory used
here is free of self interaction and the single particle orbitals for
up and down spins may have different spatial variations, allowing a
description of magnetic states and localized electrons.

\begin{figure}
\begin{center}
\includegraphics{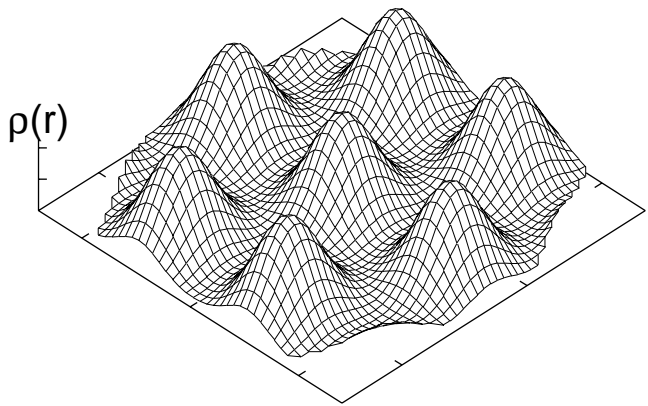}
\end{center}
\caption{\label{2d_F_rs=10_charge_density} Hartree-Fock charge density
(arbitrary units) of the 2D ferromagnetic hexagonal Wigner crystal at
$r_s = 10$.}
\end{figure}

For our 2D calculations we wrote a Hartree-Fock code which uses a
plane-wave basis set.  For our 3D calculations we used the
CRYSTAL~\cite{crystal} Gaussian basis set code.  In 2D we considered
square and hexagonal lattices with one electron per unit cell for
fully polarized systems, and two electrons per unit cell for
unpolarized systems.  We found stable Wigner crystal solutions for
$r_s \geq 1.44$.  Fig.~\ref{2d_F_rs=10_charge_density} shows the
electron density of the 2D ferromagnetic hexagonal Wigner crystal at
$r_s=10$, clearly showing the hexagonal lattice.  The ratio of the
maximum to minimum charge densities is 13 for this crystal and 17 for
the corresponding antiferromagnetic hexagonal crystal.

\begin{figure}
\begin{center}
\includegraphics{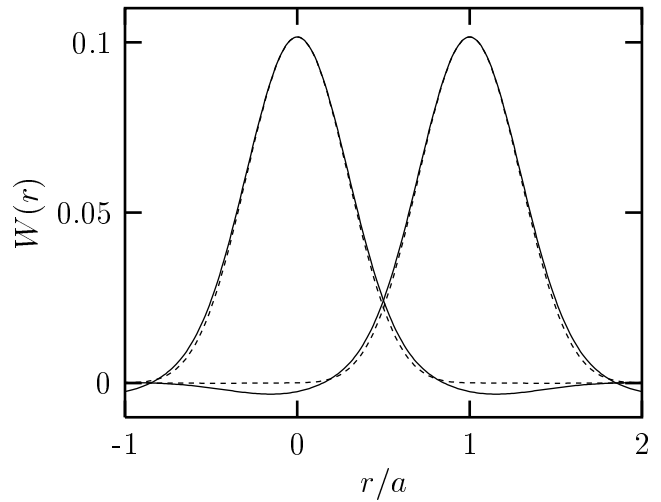}
\end{center}
\caption{\label{2d_rs=10_wannier_function} Wannier functions for the
2D antiferromagnetic hexagonal Wigner crystal at $r_s = 10$ along a
line joining two opposite-spin nearest neighbors (dashed line) and
joining two parallel-spin nearest neighbors (solid line).  The
nearest neighbor distance is $a$.}
\end{figure}

Fig.~\ref{2d_rs=10_wannier_function} shows the Wannier functions
centred on neighboring sites of the antiferromagnetic Wigner crystals
at $r_s = 10$.  The overlap of the Wannier functions is small,
indicating that the electrons are kept far apart.  Note also that the
parallel-spin Wannier functions have oscillations which maintain their
orthogonality.

\begin{figure}
\begin{center}
\includegraphics{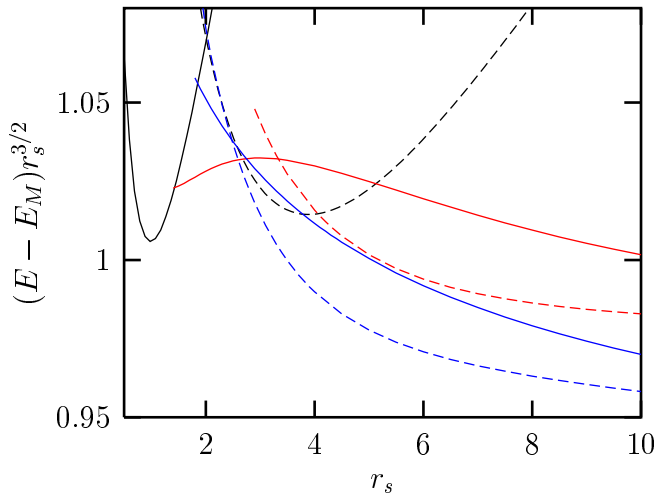}
\end{center}
\caption{\label{2d_wigner_hf_energies} Hartree-Fock energies in au per
electron of the unpolarized (solid) and fully polarized (dashed) 2D
phases as a function of $r_s$ for the square (red) and hexagonal
(blue) lattices, and for the fluid phases (black).  For clarity of
presentation we have subtracted the Madelung energy of the hexagonal
lattice, $E_M = -1.1061/r_s$, and multiplied by $r_s^{3/2}$.}
\end{figure}

Fig.~\ref{2d_wigner_hf_energies} shows the Hartree-Fock energies of
various 2D phases as a function of $r_s$.  The data shows a
first-order transition from the paramagnetic fluid phase to the
antiferromagnetic square Wigner crystal at $r_s = 1.44$ and another
first-order transition at $r_s = 2.60$ to the ferromagnetic hexagonal
crystal, which remains the most stable phase up to the highest density
studied of $r_s = 100$.  The ferromagnetic fluid phase is predicted to
be unstable at all densities.

The Hartree-Fock solutions break the translational invariance of the
many-body Hamiltonian.  There is an infinite number of degenerate
solutions corresponding to arbitrary translations and rotations, but
our calculations pick out a particular translational and rotational
state.  Using a unit cell removes the rotational degree of freedom and
for the 3D calculations the Gaussian basis set removes the
translational degeneracy.  The 2D plane-wave calculations also pick
out particular translational states, which depend on the starting
point of the iterative solution of the equations, but if we
artificially translate the final self-consistent solution we find that
the energy only changes by of order the calculational precision
($\sim$10$^{-16}$ au per electron).  Our calculations therefore lead
to symmetry broken solutions which represent ``pinned'' Wigner
crystals.  In experiments pinning of Wigner crystals may arise from
the presence of impurities or boundaries, and therefore our broken
symmetry solutions are physically meaningful.

The broken symmetry Wigner crystals described by Hartree-Fock theory
have band gaps for single electron excitations.  The ferromagnetic
crystal illustrated in Fig.~\ref{2d_F_rs=10_charge_density} has a band
gap of 2.0 eV while the corresponding antiferromagnetic crystal has a
gap of 3.3 eV.  At small $r_s$ the band gaps of the crystalline phases
rise steeply with increasing $r_s$.  On further increase of $r_s$ the
band gaps reach maximum values and then slowly decrease.

\begin{figure}
\begin{center}
\includegraphics{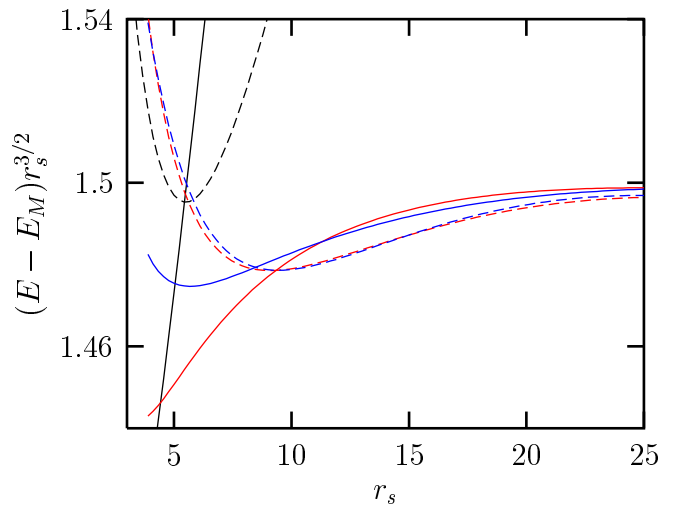}
\end{center}
\caption{\label{3d_wigner_hf_energies} Hartree-Fock energies in au per
electron of the unpolarized (solid) and fully polarized (dashed) 3D
phases as a function of $r_s$ for the bcc (red) and fcc (blue)
lattices, and for the fluid phases (black).  For clarity of
presentation we have subtracted the Madelung energy of the bcc
lattice, $E_M = -0.89593/r_s$, and multiplied by $r_s^{3/2}$.}
\end{figure}

In 3D we considered bcc and fcc lattices with one electron per unit
cell for fully polarized systems, and two electrons per unit cell for
unpolarized systems.  We found Wigner crystal solutions for $r_s \geq
4.4$.  The Hartree-Fock energies (Fig.~\ref{3d_wigner_hf_energies})
show first-order transitions from the paramagnetic fluid phase to the
antiferromagnetic bcc Wigner crystal at $r_s = 4.4$, then to the
ferromagnetic fcc Wigner crystal at $r_s = 9.5$, and finally to the
ferromagnetic bcc crystal at $r_s = 13.3$, which remains the most
stable phase up to the highest density studied of $r_s=100$.  The
ferromagnetic fluid is predicted to be unstable at all densities.  We
also found a second region ($9.5 <r_s < 9.7$) where the ferromagnetic
bcc crystal is extremely close to stability, but the resolution of our
data is insufficient to confirm whether it is actually stable in this
density range.  Note that Hartree-Fock theory predicts (incorrectly)
that the electron fluid should crystallize at the average valence
charge densities of the heavier alkali metals K, Rb and Cs.

\begin{figure}
\begin{center}
\includegraphics{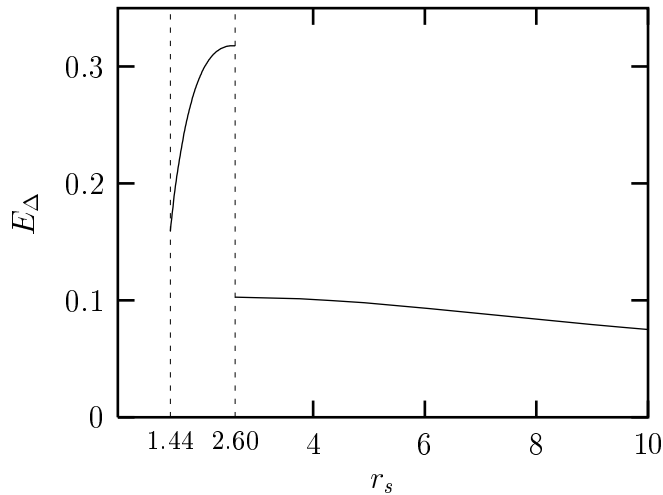}
\end{center}
\caption{\label{2d3d_correlation_energy} The ratios of the correlation
energies of the 2D hexagonal crystal and ferromagnetic fluid phases
(solid line), and of the 3D bcc crystal and ferromagnetic fluid phases
(dashed line) as a function of $r_s$.}
\end{figure}

Hartree-Fock theory gives the single determinant approximation to the
many-body wave function with the lowest possible energy, which is
always greater than (or equal to) the exact energy.  The difference
between the exact and Hartree-Fock energies, $e_{\rm c} = E_{\rm
exact} - E_{\rm HF}$, is known as the ``correlation energy''.  The
energies of the Fermi fluid and Wigner crystal phases are known
accurately from DMC
calculations~\cite{Cep&Ald,attaccalite_2002,zong_2002,rapisarda,tanatar,ortiz_1999},
and therefore we may determine the correlation energies of the
different phases.  In Fig.~\ref{2d3d_correlation_energy} we plot the
ratio $(e_{\rm c}^{\rm crystal} / e_{\rm c}^{\rm fluid})$ as a
function of $r_s$, where $e_{\rm c}^{\rm crystal}$ is the correlation
energy of the crystal phase (hexagonal in 2D and bcc in 3D) and
$e_{\rm c}^{\rm fluid}$ is the correlation energy of the ferromagnetic
fluid phase.  (The energy differences between the ferromagnetic and
antiferromagnetic crystals are negligible at these densities.)
Fig.~\ref{2d3d_correlation_energy} shows that the correlation energy
of the crystalline phase is much smaller than in the fluid in both 2D
and 3D.  Hartree-Fock theory therefore tends to favor the crystalline
phases, which it describes more faithfully than the fluid phases.

The basic mechanism for electron crystallization within Hartree-Fock
theory is that proposed by Wigner, i.e., at sufficiently low densities
crystallization greatly reduces the interaction energy with only a
small increase in the kinetic energy.  The problem can be analyzed
more deeply in terms of the Hartree and exchange terms provided by our
calculations.  In Hartree-Fock theory one normally defines the Hartree
energy and potential to include the unphysical self-interaction,
which, however, exactly cancels the self-exchange.  In Wigner crystals
the self-interaction terms are very large and therefore it is more
illuminating to discuss the Hartree and exchange terms with the
unphysical self-interactions removed.  From this viewpoint the
essential physics of the Wigner crystal is that the electrons are kept
apart by the Hartree potential.  Exchange effects are small in Wigner
crystals at low densities.  The single particle orbitals obtained at
the Hartree-Fock level for a Wigner crystal already keep the electrons
well separated, and therefore their correlation energies are small.

As Wigner argued~\cite{wigner}, at low densities the kinetic energy is
unimportant and the many-body wave function of a Wigner crystal is
expected to have a large weight for configurations in which the
electrons lie far apart on a lattice.  The lattice adopted should
therefore have the lowest Madelung electrostatic energy, i.e., the
hexagonal lattice in 2D and the bcc lattice in 3D.  At higher
densities the kinetic energy becomes important in determining the
structure of the crystal.  There are two factors which control the
kinetic energy of Wigner crystals.  First, crystal structures with
larger packing fractions have lower kinetic energies because they
allow the electron orbitals to spread out over a greater volume (or
area in 2D) without overlap.  Second, the kinetic energy of an
antiferromagnet tends to be lower than that of the corresponding
ferromagnet because in the antiferromagnet the Wannier functions on
neighboring sites need not be orthogonal and therefore they can
overlap without oscillation, which reduces the kinetic energy.

The relative stabilities of the phases are controlled by the
competition between the kinetic and potential energy terms.  A
detailed study of the numerical values of these terms reveals the
following simple picture.  In both 2D and 3D the low-density stable
phase has the structure with the lowest Madelung energy, i.e., the
hexagonal and bcc phases, respectively.  The dominant effect at low
densities is therefore the Madelung energy, as proposed by Wigner, and
ferromagnetism is slightly favored because of the larger exchange
interactions.  The kinetic energy becomes more important at higher
densities and lattices with higher packing fractions are favored.  In
2D the hexagonal crystal has the largest packing fraction and so the
stable phase remains unchanged, but in 3D the fcc crystal has the
largest packing fraction, and therefore the ferromagnetic fcc crystal
becomes the most stable.  At still higher densities the reduction in
kinetic energy arising from adopting an antiferromagnetic spin
configuration dominates.  In 2D the hexagonal lattice frustrates
antiferromagnetism and therefore an unfrustrated square lattice
becomes more stable.  The 2D square lattice with ferromagnetic order
along the rows in one direction but antiferromagnetic order in the
perpendicular direction is calculated to have a substantially higher
energy than the completely antiferromagnetic lattice at high
densities, indicating the importance of the spin ordering.  The 3D fcc
lattice frustrates antiferromagnetism and therefore the
antiferromagnetic bcc lattice becomes more stable at higher densities.
At the very highest densities the need to reduce the kinetic energy
becomes paramount and the crystalline phases become unstable to the
formation of paramagnetic fluids.  This simple picture explains the
occurrence of the different stable phases as a function of density.

It is important to understand how different levels of theory describe
such an important model system as electrons in a uniform potential.
However, there are, of course, important corrections to Hartree-Fock
theory due to electron correlation.  The main effect of adding
correlation is to lower the energies of the fluid phases more than the
crystalline ones, which moves the transitions to crystalline phases to
lower densities.  DMC calculations show that only the 2D hexagonal and
3D bcc Wigner crystals are stable when electron correlation is
included.  In both 2D and 3D Hartree-Fock theory predicts the same
stable low density phases as DMC, which is a further indication that
Hartree-Fock theory provides a simple and useful framework for
understanding Wigner crystals.

In conclusion, we have shown that unrestricted Hartree-Fock theory is
able to describe Wigner crystals in 2D and 3D.  We believe this to be
important for four reasons.  (1) It leads to a picture of Wigner
crystals as phases with small correlation energies.  (2) It gives
simple physical insights into the competition between kinetic and
potential energy terms which determines the stability of different
phases.  (3) Hartree-Fock theory is fairly accurate and
computationally inexpensive and therefore may be used to describe
Wigner crystals in more complicated situations, such as when defects
or external fields are present or when atomistic effects are
important.  (4) Hartree-Fock theory forms a natural starting point for
more accurate descriptions of Wigner crystals, such as perturbation
theory.

We thank Gavin Brown for useful discussions.  Financial support was
provided by EPSRC (UK).


\begin{thebibliography}{}
\bibitem{density} In 2D $\pi r_s^2 = A/N$, while in 3D $4\pi r_s^3/3 =
\Omega/N$, where $N/A$ is the number of electrons per unit area, and
$N/\Omega$ is the number per unit volume.
\end{thebibliography}
\end{document}